 \documentclass[doublespacing]{elsart}

 \usepackage{graphicx}
\usepackage{amssymb}

\newcommand{\ie}{i.e.}

\begin{document}

\begin{frontmatter}



%\title{Unbound excited states in ${\bm ^{19,17}{\rm C}}$}
\title{Unbound excited states in \nuc{19,17}{C}}


\author[titech]{Y.~Satou},
\ead{satou@phys.titech.ac.jp}
\author[titech]{T.~Nakamura},
\author[riken]{N.~Fukuda},
\author[riken]{T.~Sugimoto},
\author[riken]{Y.~Kondo},
\author[titech]{N.~Matsui},
\author[titech]{Y.~Hashimoto},
\author[titech]{T.~Nakabayashi},
\author[titech]{T.~Okumura},
\author[titech]{M.~Shinohara},
\author[riken]{T.~Motobayashi},
\author[riken]{Y.~Yanagisawa},
\author[riken]{N.~Aoi},
\author[riken]{S.~Takeuchi},
\author[riken]{T.~Gomi},
\author[rikkyo]{Y.~Togano},
\author[rikkyo]{S.~Kawai},
\author[riken]{H.~Sakurai},
\author[tokyo]{H.~J.~Ong},
\author[tokyo]{T.~K.~Onishi},
\author[cns]{S.~Shimoura},
\author[cns]{M.~Tamaki},
\author[tohoku]{T.~Kobayashi},
\author[riken]{H.~Otsu},
\author[tohoku]{Y.~Matsuda},
\author[tohoku]{N.~Endo},
\author[tohoku]{M.~Kitayama}, \\
and \author[riken]{M.~Ishihara}

\address[titech]{Department of Physics, Tokyo Institute of Technology, 
2-12-1 Oh-Okayama, Meguro, Tokyo 152-8551, Japan}
\address[riken]{The Institute of Physical and Chemical Research (RIKEN), 
2-1 Hirosawa, Wako, Saitama 351-0198, Japan}
\address[rikkyo]{Department of Physics, Rikkyo University, 
3 Nishi-Ikebukuro, Toshima, Tokyo 171-8501, Japan}
\address[tokyo]{Department of Physics, University of Tokyo, 
7-3-1 Hongo, Bunkyo, Tokyo 113-0033, Japan}
\address[cns]{Center for Nuclear Study (CNS), University of Tokyo, 
2-1 Hirosawa, Wako, Saitama 351-0198, Japan}
\address[tohoku]{Department of Physics, Tohoku University, 
Aoba, Sendai, Miyagi 980-8578, Japan}

\begin{abstract}
% Text of abstract
The neutron-rich carbon isotopes $^{19,17}{\rm C}$ 
have been investigated via proton inelastic scattering 
on a liquid hydrogen target at 70 MeV/nucleon. 
The invariant mass method in inverse kinematics was employed 
to reconstruct 
%extract 
the energy spectrum, 
in which fast neutrons and charged fragments 
were detected in coincidence using a neutron hodoscope 
and a dipole magnet system. 
A peak has been observed 
with an excitation energy of 1.46(10) MeV in $^{19}{\rm C}$, 
while three peaks with energies of 2.20(3), 3.05(3), and 6.13(9) MeV 
have been observed in $^{17}{\rm C}$. 
%The 
Deduced cross sections 
%leading to the 1.46 and 2.20 MeV states 
are compared with microscopic DWBA calculations 
based on $p$-$sd$ shell model wave functions 
and modern nucleon-nucleus optical potentials. 
$J^{\pi}$ assignments are made 
for the four observed states as well as the ground states of both nuclei. 
\end{abstract}

\end{frontmatter}

% main text
%\section{}
%\label{}

With the advent of new radioactive beam facilities 
capable of producing intense beams of various nuclear species 
far from the stability line, even at the drip-line, 
an increasingly large amount of 
%more and more 
information on nuclear levels and modes of excitation 
is being accumulated throughout the nuclear chart. 
New phenomena, such as nuclear halos and skins~\cite{Tanihata85}, 
enhanced $E1$ transition strengths~\cite{Nakamura94-06,Adrich05}, 
and 
modifications of shell closures~\cite{Iwasaki00,Ozawa00} 
have been revealed. 
Of particular interest in recent years are the neutron-rich carbon isotopes, 
which have attracted attention not only from their own structural interest, 
such as the anomalously reduced $E2$ transition strength in 
$^{16}{\rm C}$~\cite{Imai04,Elekes04}, 
but also 
for their implications for 
%from an implication to 
a new magic number at the neutron number $N$=16 
proposed in oxygen isotopes~\cite{Ozawa00,Elekes07}. 

Neutron number dependence of ground state deformations of carbon isotopes 
has been investigated in a deformed Hartree-Fock (HF) theory~\cite{Suzuki03a}. 
Generally the prolate deformation is expected at the beginning of the shell 
whereas oblate deformation arises towards the end of the shell. 
Note that the occurrence of nuclear deformations in the ground state 
is a consequence of the spontaneous symmetry breaking effect 
known in many fields of physics, 
and there is broad 
%a wide 
interest 
in elucidating the intriguing deformation-driving 
mechanism in atomic nuclei~\cite{Sagawa04}, 
which would be a nuclear physics analogue 
of the Jahn-Teller effect in molecular physics~\cite{Jahn37}. 
The theory predicts prolate deformations for carbon isotopes with $N$=9--11. 
For $^{19}{\rm C}$ with $N$=13 
two almost degenerate deformed minima 
%minima almost degenerated in energy 
are predicted 
%at both sides 
with 
%different 
spins 
$J^{\pi}$=$1/2^+$ (prolate) 
%for prolate 
and $3/2^+$ (oblate). 
%for oblate deformations. 
Since the shape change at a neutron number 
smaller than the middle of $N$=8 and 20 
indicates a new shell closure at $N$=16, 
it is argued that definite information on the structure of $^{19}{\rm C}$ 
is important to clarify the possible new shell effect 
in the neutron-rich carbon isotopes~\cite{Suzuki03a,Suzuki03b}. 

The $^{19}{\rm C}$ nucleus is the heaviest odd carbon isotope. 
It is loosely bound with a neutron separation energy 
of $S_n$=0.58(9) MeV~\cite{Audi03}, 
and exhibits one-neutron halo structure as evidenced 
by large Coulomb break-up cross sections~\cite{Nakamura99}. 
The ground state spin and parity were 
%property was 
also investigated via the measurements 
of longitudinal momentum distributions of charged fragments 
after the removal of one neutron 
from $^{19}{\rm C}$~\cite{Bazin95,Maddalena01}. 
From these measurements 
the spin-parity was assigned to be $J^{\pi}_{\rm g.s.}$=$1/2^+$. 
Other possibilities, however, 
have been suggested by authors in Ref.~\cite{Kanungo05} 
based on their experiment 
in search of an isomeric transition. 
Moreover, there remains a controversy over the different widths 
of longitudinal momentum distributions measured 
at different energies~\cite{Bazin98,Baumann98}, 
which has led to a conjecture of a possible resonance state 
just above the particle decay threshold 
as a clue to solve such an inconsistency~\cite{Smedberg99}. 

In this situation 
it it worthwhile accumulating experimental information 
on the ground as well as excited states of $^{19}{\rm C}$. 
This paper reports a new measurement 
using the $(p,p')$ inelastic scattering on $^{19}{\rm C}$ 
employing the invariant mass method in inverse kinematics. 
Both decaying neutrons and charged fragments were detected, 
and an isolated level was 
identified 
%selected 
in the final state. 
The measurement was also made, 
partly for calibration purposes, 
on another loosely bound nucleus $^{17}{\rm C}$, 
having $S_n$=0.73(2) MeV~\cite{Audi03}, 
for which $J^{\pi}_{\rm g.s.}$ is consistently reported 
to be $3/2^+$~\cite{Maddalena01,Ogawa02,Sauvan00-04,Pramanik03}. 

The $(p,p')$ reaction has the following advantages: 
(1) The magnitude of the cross section, 
%The cross section magnitude, 
being sensitive to both the initial and final state wave functions, 
is configuration dependent. 
(2) The shape of the angular distribution 
%The angular distribution shape 
also depends on the configuration 
through characteristic transferred $L$ dependences of partial amplitudes. 
(3) Theoretical methods such as the distorted-wave Born approximation (DWBA) 
can be used to provide a first interpretation of data. 
%applies in interpreting data. 
There exists one $(p,p')$ work on $^{19}{\rm C}$ 
reporting two bound states at 0.20 and 0.27 MeV 
using $\gamma$-ray spectroscopy~\cite{Elekes05}. 
These states are tentatively assigned as $3/2^+$ and $5/2^+$, respectively, 
based on the assumption of $J^{\pi}_{\rm g.s.}$=$1/2^+$. 
Close proximity of levels near the ground state 
has made it difficult to identify levels from comparisons in excitation energy 
between theory and experiment. 
Note that 
in shell model calculations~\cite{Maddalena01} 
the triplet of levels 
%with $J^{\pi}$=$1/2^+$, $3/2^+$, and $5/2^+$ 
are predicted below 0.62 MeV 
with no 
%unique 
ground state configuration favoured. 
We have chosen to probe states in the unbound region, 
where in contrast to the bound region a lower level density is predicted up to 
about 3 MeV from the threshold in a shell model calculation described later. 
With no states known above the particle decay threshold, 
the measurement involved a search for resonances in this region, 
and we report a new state in this paper. 
For $^{17}{\rm C}$ eleven new states up to 16.3 MeV excitation energy 
have been recently 
reported from 
%identified via 
the three-neutron transfer reaction 
$^{14}{\rm C}$($^{12}{\rm C}$,$^9{\rm C}$)$^{17}{\rm C}$, 
with limited spin assignments~\cite{Bohlen07}. 
This nucleus is also known to have a low-lying triplet 
of levels $1/2^+$, $3/2^+$, and $5/2^+$ below 0.33 MeV~\cite{Elekes05}. 

The experiment was performed 
at the RIKEN Accelerator Research Facility (RARF). 
The radioactive beams of $^{19}{\rm C}$ and $^{17}{\rm C}$ 
at 70 MeV/nucleon 
were produced using the projectile-fragment separator, RIPS~\cite{Kubo92}, 
from a $^{22}{\rm Ne}$ primary beam at 110 MeV/nucleon. 
Typical beam intensities were 260 cps for $^{19}{\rm C}$ 
and 10.4 kcps for $^{17}{\rm C}$ 
with momentum spreads $\Delta P/P$ of 3.0\% and 0.1\%, respectively. 
The secondary target was a cryogenic hydrogen target~\cite{Ryuto06} 
having a cylindrical shape, 
3 cm in diameter and 120$\pm$2 mg/cm$^2$ in thickness. 
%The thickness 
%%of the target 
%was chosen not to deteriorate the invariant mass 
%resolution but to obtain reasonable counting statistics. 
%Ejected 
The target was surrounded by forty-eight NaI(Tl) scintillators 
used to detect de-excitation $\gamma$-rays from the charged fragments. 
Each crystal had a dimension of 6.1$\times$6.1$\times$12.2 cm$^3$. 
The charged fragments 
%particles 
were bent by a dipole magnet behind the target 
and were detected by a plastic counter hodoscope. 
Multi-wire drift chambers placed before and after the magnet 
were used to extract trajectory information of the charged particles. 
%determine the trajectory of the charged particles. 
Neutrons were detected by a neutron hodoscope 
%consisted 
consisting of two walls of a plastic scintillator array 
placed 4.6 and 5.8 m behind the target. 
Each wall had a dimension of 2.14$^{\rm W}$$\times$0.72$^{\rm H}$
(or 0.90$^{\rm H}$)$\times$0.12$^{\rm T}$ m$^3$. 
%two 
%plastic scintillator hodoscopes 
%%neutron counter hodoscopes 
%placed 4.6 m and 5.8 m behind the target; 
%each had a dimension of 
%2.14$^{\rm W}$$\times$0.72$^{\rm W}$(or 0.90$^{\rm H}$)$\times$0.12$^{\rm T}$ 
%m$^3$. 
%Total detection efficiency of the hodoscope 
The total efficiency of the hodoscope 
was 
24.1$\pm$0.8\% 
%24.1$\pm$0.3\% 
for a threshold setting of 4 MeVee. 
This was 
deduced by measuring the $^7{\rm Li}(p,n)^7{\rm Be}$(g.s.+0.43 MeV) reaction 
at $E_p$=70 MeV and using existing cross section data~\cite{Taddeucci90}. 
%measured to be 24.1$\pm$0.3\% at a threshold setting of 4 MeVee 
%by using the $^7{\rm Li}(p,n)^7{\rm Be}$(g.s.+0.43 MeV) reaction 
%at $E_p$=70 MeV 
%and the known cross sections~\cite{Taddeucci90}. 
The invariant mass of the final system was calculated event-by-event 
from the momentum vectors of the charged particle and the neutron. 
A series of studies 
probing unbound resonance states in beryllium isotopes 
has been successfully performed 
using a detector setup similar to the present 
one~\cite{Fukuda04,Sugimoto06,Kondo07}. 

\begin{figure}[P]
\begin{center}
\includegraphics*[width=18cm,angle=-90]{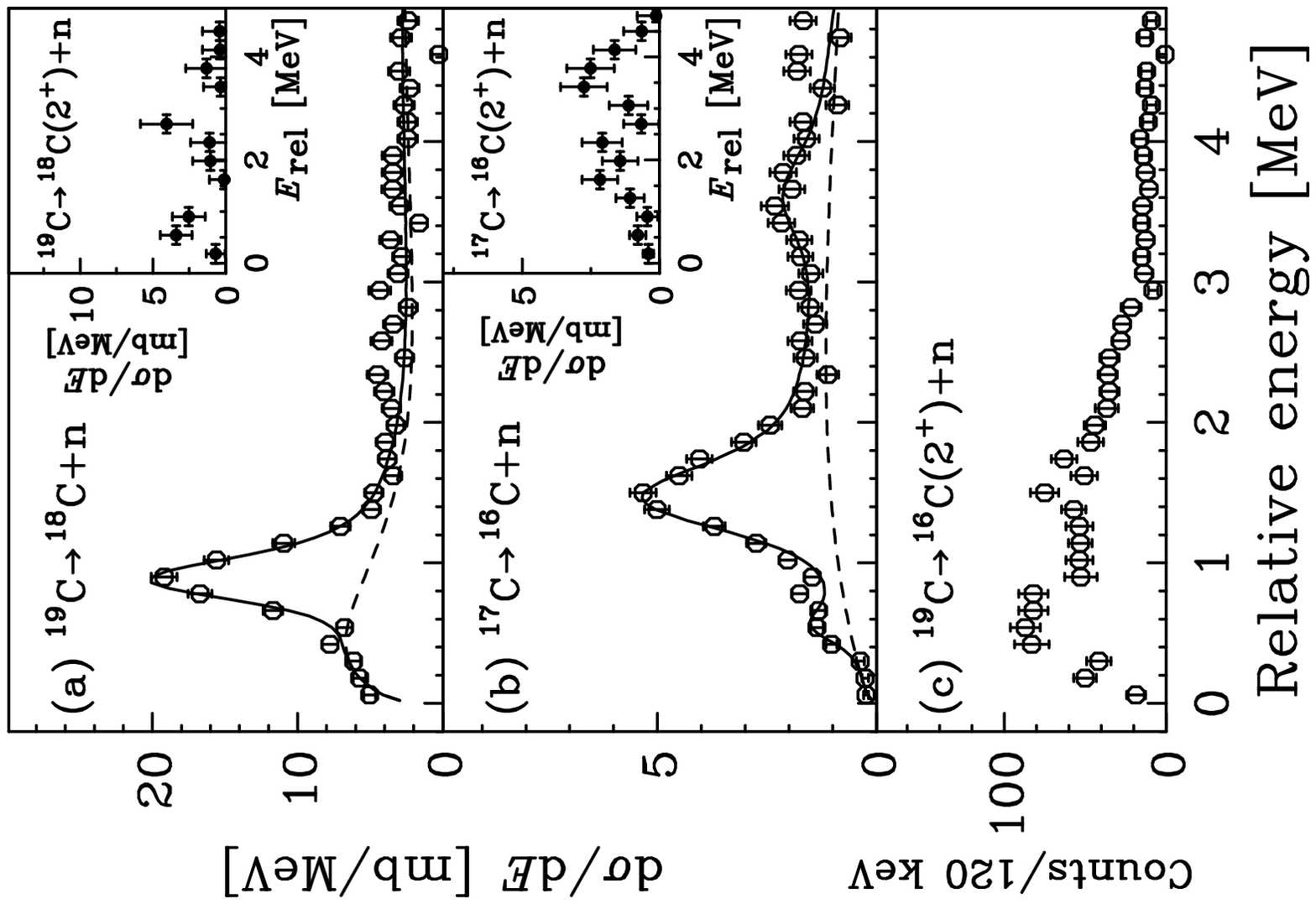}
\end{center}
\caption{Relative energy spectra 
integrated over an angular range below $\theta_{\rm c.m.}$=64$^{\circ}$ 
for the (a) $^1{\rm H}(^{19}{\rm C},$$^{18}{\rm C}$+$n)$, 
(b) $^1{\rm H}(^{17}{\rm C},$$^{16}{\rm C}$+$n)$, and 
(c) $^1{\rm H}(^{19}{\rm C},$$^{16}{\rm C}(2^+;1.77$ MeV)+$n)$ reactions 
at 70 MeV/nucleon. 
Shown in the insets of panels (a) and (b) are spectra 
obtained in coincidence with de-excitation $\gamma$-rays from the first 
$2^+$ states of the respective charged fragments. 
The instrumental background was subtracted in the spectra. 
Solid lines represent the results of the fit, 
dashed lines background introduced to reproduce the overall spectrum. }
\label{fig:spectrum_fit_gamma_plb}
\end{figure}

Figure~\ref{fig:spectrum_fit_gamma_plb} shows relative energy spectra 
for the (a) $^1{\rm H}$($^{19}{\rm C},$$^{18}{\rm C}$+$n)$, 
(b) $^1{\rm H}$($^{17}{\rm C},$$^{16}{\rm C}$+$n)$, 
and (c) $^1{\rm H}$($^{19}{\rm C},$$^{16}{\rm C}(2^+;1.77$ MeV)+$n)$ 
reactions, 
integrated over center-of-mass angles up to $\theta_{\rm c.m.}$=$64^{\circ}$.
The effect of the finite detector acceptance was corrected for 
in panels (a) and (b), \ie{} in inelastic channels. 
Shown in the insets of panels (a) and (b) are spectra 
%in the same inelastic channels, 
obtained in coincidence with de-excitation $\gamma$-rays 
from the first $2^+$ states at 1.58(1) MeV in $^{18}{\rm C}$~\cite{Stanoiu04} 
and at 1.77(1) MeV in $^{16}{\rm C}$~\cite{Tilley93}, respectively. 
The spectrum in panel (c) also required the coincidence detection 
of $\gamma$-rays from the $2^+$ state in $^{16}{\rm C}$. 
Background contributions from various window materials surrounding the target, 
measured with an empty target, 
are subtracted. 
Error bars represent statistical uncertainties. 
Peak structures at about 0.9 MeV in relative energy in $^{19}{\rm C}$ 
and about 1.5 MeV in $^{17}{\rm C}$ 
are clearly seen in spectra without requiring $\gamma$-ray coincidences. 
They are absent in the respective $\gamma$-ray coincidence spectra. 
We thus conclude that charged particles resulting from the decay of 
these peaks are in their ground states. 
In panel (b) 
we can see two more peaks 
at about 0.6 and 3.6 MeV in relative energy. 
These are visible in the $\gamma$-ray coincidence spectrum 
in the inset of panel (b); 
the 0.6 MeV peak is more clearly populated in the spectrum in panel (c). 
This observation indicates that these peaks in $^{17}{\rm C}$ 
decay through the first $2^+$ state in $^{16}{\rm C}$. 

The experimental spectra were analyzed 
to extract the resonance energy $E_r$ and the width $\Gamma_r$ 
in the following way. 
Firstly, 
a single Breit-Wigner shape function was generated 
for certain values of $E_r$ and $\Gamma_r$: 
\begin{equation}
\sigma(E_{\rm rel})\sim\frac{\Gamma_l(E_{\rm rel})\Gamma_r}
{\{E_r+\Delta E_l(E_{\rm rel})-E_{\rm rel}\}^2
+\{\Gamma_l(E_{\rm rel})/2\}^2}. 
\end{equation}
The shift function $\Delta E_l(E_{\rm rel})$ 
and the level width $\Gamma_l(E_{\rm rel})$, 
which depend on the relative energy $E_{\rm rel}$, 
were calculated by using the penetration $P_l$ and 
shift $S_l$ factors~\cite{Lane-Thomas58} 
by the relations: 
\begin{eqnarray}
\Delta E_l(E_{\rm rel})&=&
\Gamma_r\times \{S_l(E_r)-S_l(E_{\rm rel})\}/\{2P_l(E_r)\}, \\
\Gamma_l(E_{\rm rel})&=&
\Gamma_r\times P_l(E_{\rm rel})/P_l(E_r), 
\end{eqnarray}
where $l$ refers to the decay angular momentum. 
Decay neutrons were supposed to be in the $l$=2 orbit. 
%$d$-orbit. 
The channel radius was taken to be 
$R$=$r_0(A_n^{1/3}+A_f^{1/3})$ with $r_0$=1.4 fm, 
where $A_n$ and $A_f$ 
are the neutron and charged fragment mass numbers, 
respectively. 
Then, 
the experimental resolution, 
including the detector resolution, beam profile, 
Coulomb multiple scattering of charged particles, 
and their range difference in the secondary target, 
was incorporated to generate a response function. 
The relative energy resolution was well simulated 
by $\Delta E_{\rm rel}$=0.13$\sqrt{E_{\rm rel}}$ MeV (rms). 
The procedure was repeated for each resonance peak by varying 
the initial values of $E_r$ and $\Gamma_r$. 
Parameters which gave the best fit to the data were obtained 
by minimizing the $\chi^2$. 
%finding $\chi^2$ minimum. 
The solid curves in Fig.~\ref{fig:spectrum_fit_gamma_plb} 
show the results of the fit; 
the dashed lines background introduced 
to reproduce the overall spectrum. 
%%final response functions on top of the background 
%%indicated by 
%%shown as 
%%dashed lines. 
%%the shape of which was appropriately chosen 
%%to reproduce the overall spectrum. 
%%The functional shape of the latter was appropriately chosen 
%%to reproduce the overall spectrum. 
%%show the results of the fit. 
%%The dashed lines represent the background; 
%%the functional shape was appropriately chosen 
%%to reproduce the overall spectrum. 

\begin{figure}[P]
\begin{center}
\includegraphics*[height=11cm,angle=-90]{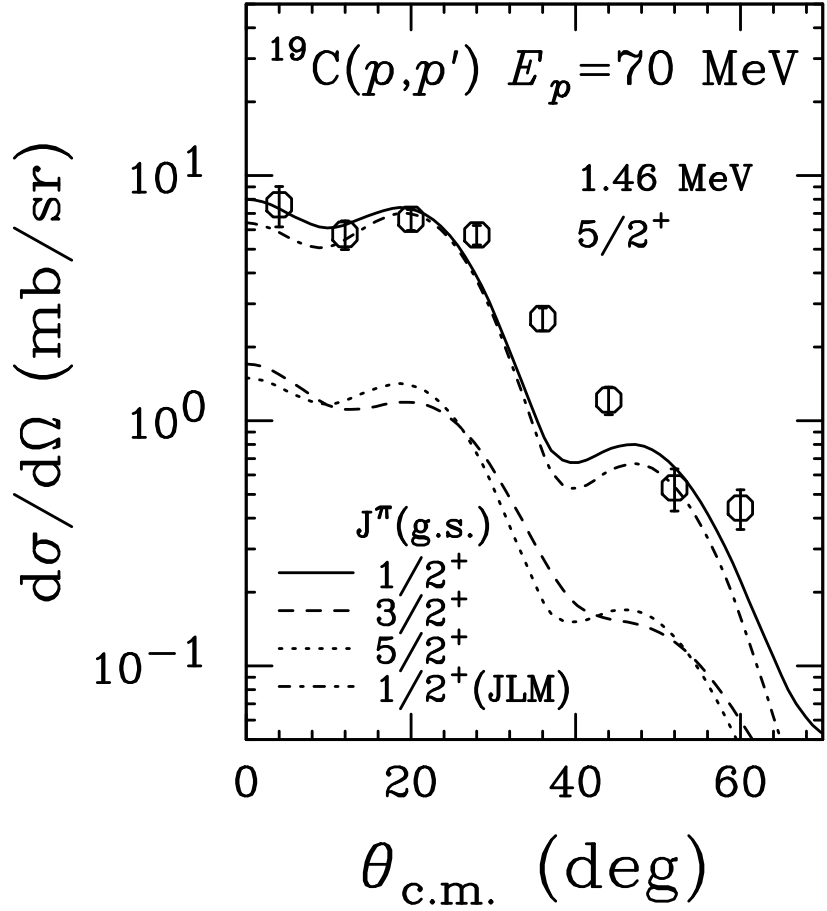}
\end{center}
\caption{Differential cross sections leading 
to the 1.46 MeV state in $^{19}{\rm C}$ 
are compared to DWBA predictions obtained 
by assuming different initial-state configurations. 
$J^{\pi}$=$5/2^+_2$ is assumed for the excited state. }
\label{fig:neut_19c_ad_plb}
\end{figure}

\begin{figure}[P]
\begin{center}
\includegraphics*[height=11cm,angle=-90]{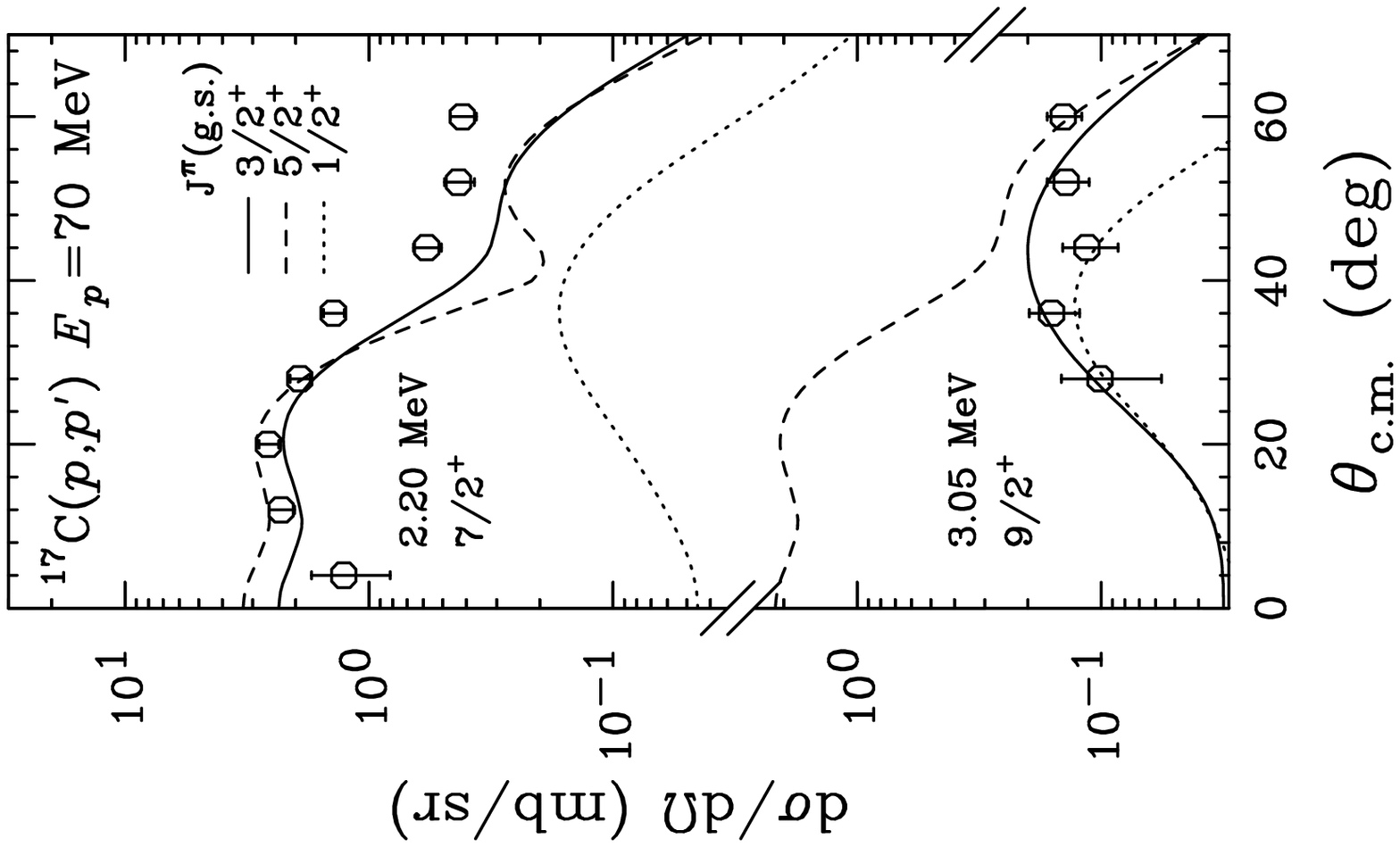}
\end{center}
\caption{Differential cross sections leading 
to the 2.20 and 3.05 MeV states in $^{17}{\rm C}$ 
are compared to DWBA predictions obtained 
by assuming different initial-state configurations. 
$J^{\pi}$ values assumed for the excited states 
are $7/2^+_1$ and $9/2^+_1$ for the 2.20 and 3.05 MeV states, 
respectively. }
\label{fig:neut_17c_ad_plb}
\end{figure}

Resonance parameters extracted 
are summarized in Table~\ref{tbl:resonance_parameters}. 
%Uncertainties in $E_r$ and $\Gamma_r$ 
%are taken to be the amount of variations from the best fit values, 
%which leads to an increase in $\chi^2$ by 1. 
The excitation energy $E_{\rm x}$ 
is obtained by the relation: $E_{\rm x}$=$E_r$+$S_n$+$E^{*}$, 
where $E^{*}$ refers to the excitation energy of the daughter nucleus. 
The state at $E_{\rm x}$=1.46(10) MeV in $^{19}{\rm C}$ 
was observed for the first time in this measurement. 
The 2.20(3), 3.05(3), and 6.13(9) MeV states in $^{17}{\rm C}$ 
are respectively close to the 2.06, 3.10, and 6.20 MeV states 
observed in the three-neutron transfer reaction~\cite{Bohlen07}. 
Extracted widths for the 1.46 and 2.20 MeV states 
are $\Gamma_r$=0.29(2) and 0.53(4) MeV, respectively. 
These values are of the order of 
%comparable to 
the respective single-particle estimates for the width 
of $l$=2 resonances~\cite{Bohr68}, 
$\Gamma_{\rm sp}$=0.25 and 0.78 MeV, 
supporting the assumption of $l$=2 decays of these states. 
%that $l$=2 neutrons are responsible for the decay. 
For the 3.05 MeV state 
the $\Gamma_r$ value was not extracted 
since it was insensitive to the spectrum shape 
owing to the 
%due to 
very small penetration factors at low relative energies. 
Cross sections leading to these states, 
angle-integrated up to $\theta_{\rm c.m.}$=64$^{\circ}$, 
are given in Table~\ref{tbl:resonance_parameters}. 
Differential cross sections leading to the 1.46 MeV state in $^{19}{\rm C}$ 
are shown in Fig.~\ref{fig:neut_19c_ad_plb}, 
and those to the 2.20 and 3.05 MeV states in $^{17}{\rm C}$ 
are shown in Fig.~\ref{fig:neut_17c_ad_plb}. 
The errors shown 
are the quadratic sum of statistical and systematic uncertainties; 
the latter in the absolute magnitude of the cross section 
is estimated to be 7\%, 
including ambiguities in target thickness, 
neutron detection efficiency, 
and fitting procedure. 
The angular resolution varied 
from $\Delta\theta_{\rm c.m.}$=3.4$^{\circ}$ (3.1$^{\circ}$) in rms 
at $\theta_{\rm c.m.}$=4$^{\circ}$ 
to 5.1$^{\circ}$ (4.6$^{\circ}$) at 60$^{\circ}$ 
for $^{19}{\rm C}$ ($^{17}{\rm C}$). 
%This was estimated using a Monte Carlo simulation 
%which took into account the detector resolution, 
%the Coulomb multiple scattering of charged particles, 
%and the range difference of them in the target. 
%supporting the assumption 
%on the orbital angular momentum of the decaying neutron ($d$-orbit) 
%in the response function analysis. 
%suggesting that the major contribution to the width 
%comes from the decay of a d-wave neutron. 

To obtain an understanding of the results 
shell model calculations were performed using the code OXBASH~\cite{OXBASH88} 
within the 0$\hbar\omega$ configurations in the $p$-$sd$ shell model space. 
For $^{19}{\rm C}$, 
the second $5/2^+_2$ state is predicted at 1.40, 1.54, and 1.47 MeV 
with the WBT, WBP~\cite{Warburton92}, and MK~\cite{Millener75} 
interactions, respectively. 
These energies are close to the experimental value of 1.46 MeV; 
it is likely that this state corresponds to the $5/2^+_2$ state. 
%leading to the $5/2^+_2$ assignment for this state. 
%reasonably close 
%%to each other and 
%to the energy of 1.46 MeV of the observed peak. 
%The 1.46 MeV state is thus likely to be the $5/2^+_2$ state. 
%The good agreement in energy allows the $5/2^+_2$ assignment for this state. 
%will lead to the assignment of $5/2^+_2$ for this state. 
%the state at 1.46 MeV. 
%These 
%are very near to the observed excitation energy of $E_{\rm x}$=1.46 MeV; 
%it is likely that this peak has a $J^{\pi}$ value of $\frac52^+$. 
%The next state predicted is the first $7/2^+$ state, 
%which is 700 keV above the $5/2^+_2$ state with the PSDWBT interaction. 
The closest higher-lying state is the $7/2^+_1$ state 
predicted 0.7--1.0 MeV above the $5/2^+_2$ state. 
%The next state predicted is the $7/2^+_1$ state 
%lying 0.7--1.0 MeV above the $5/2^+_2$ state. 
For $^{17}{\rm C}$, 
with the WBT interaction 
four states $5/2^+_2$(1.72 MeV), $7/2^+_1$(2.33 MeV), 
$9/2^+_1$(3.01 MeV), and $3/2^+_2$(3.08 MeV) 
are predicted 
above the decay threshold and below 3.5 MeV. 
%in the relevant energy region. 
%above the threshold and 
%below 3.5 MeV.

\begin{table}[p]
\begin{center}
\caption{Resonance parameters and populating cross sections 
from this experiment. 
The excitation energy and the cross section 
are compared to theoretical values. 
Cross sections are integrated up to $\theta_{\rm c.m.}$=64$^{\circ}$. 
DWBA cross sections $\sigma_{\rm DWBA}$ 
leading to states specified by $J^{\pi}$ 
are obtained by using the shell model wave functions 
with the WBT interaction, 
assuming $J^{\pi}_{\rm g.s.}$=1/2$^+_1$ for $^{19}{\rm C}$ 
and 3/2$^+_1$ for $^{17}{\rm C}$. 
\label{tbl:resonance_parameters}} 
\begin{tabular}{crrrrrrrr} \hline \hline
 & 
\multicolumn{4}{c}{Experiment} & &
\multicolumn{3}{c}{Theory} \\ 
\cline{2-5}
\cline{7-9}
Nucleus & 
\multicolumn{1}{c}{$E_r$} & 
\multicolumn{1}{c}{$E_{\rm x}$} & 
\multicolumn{1}{c}{$\Gamma_r$} & 
\multicolumn{1}{c}{$\sigma_{\rm exp}$} & & 
\multicolumn{1}{c}{$E_{\rm x}$} & 
\multicolumn{1}{c}{$\sigma_{\rm DWBA}$} & 
\multicolumn{1}{c}{$J^{\pi}$} \\ 
  &
\multicolumn{1}{c}{(MeV)} & 
\multicolumn{1}{c}{(MeV)} & 
\multicolumn{1}{c}{(MeV)} & 
\multicolumn{1}{c}{(mb)}  & &
\multicolumn{1}{c}{(MeV)} & 
\multicolumn{1}{c}{(mb)}  &
                              \\ \hline
$^{19}{\rm C}$ & 
\multicolumn{1}{c}{0.88(1)} & 
\multicolumn{1}{c}{\hspace*{1.1ex}1.46(10)} & 
\multicolumn{1}{c}{0.29(2)} & 
\multicolumn{1}{c}{8.6(4)} & &
\multicolumn{1}{c}{1.40} & 
\multicolumn{1}{c}{6.9\hspace*{1.2ex}} & 
\multicolumn{1}{c}{5/2$^+_2$}  \\ 
$^{17}{\rm C}$ & 
\multicolumn{1}{c}{1.47(2)} & 
\multicolumn{1}{c}{2.20(3)} & 
\multicolumn{1}{c}{0.53(4)} & 
\multicolumn{1}{c}{3.8(2)} & &
\multicolumn{1}{c}{2.33} & 
\multicolumn{1}{c}{2.7\hspace*{1.2ex}} & 
\multicolumn{1}{c}{7/2$^+_1$}  \\ 
               & 
\multicolumn{1}{c}{0.55(2)} & 
\multicolumn{1}{c}{3.05(3)}  & 
\multicolumn{1}{c}{---} & 
\multicolumn{1}{c}{\hspace*{1.1ex}0.40(4)} & &
\multicolumn{1}{c}{3.01} & 
\multicolumn{1}{c}{0.48}  &
\multicolumn{1}{c}{9/2$^+_1$}  \\
               & 
\multicolumn{1}{c}{3.63(9)} & 
\multicolumn{1}{c}{6.13(9)} & 
\multicolumn{1}{c}{\hspace*{1.7ex}0.26$^{+0.4}_{-0.26}$} & 
\multicolumn{1}{c}{0.8(1)} & &
\multicolumn{1}{c}{6.25} & 
\multicolumn{1}{c}{0.94} &
\multicolumn{1}{c}{5/2$^+_4$} \\
\hline
\hline
\end{tabular}
\end{center}
\end{table}

In order to further clarify the nature of states 
%identify 
%transitions 
%initial and final states 
%configurations 
involved in transitions shown in 
Table~\ref{tbl:resonance_parameters}, 
%Fig.~\ref{fig:neut_19c_17c_ad_plb}, 
microscopic DWBA calculations 
were performed using the code DW81~\cite{Raynal70}. 
The optical potential was taken from the global parameterization 
KD02~\cite{Koning03}. 
A microscopic optical potential~\cite{Garcia05} 
based on the approach of Jeukenne, Lejeune, and Mahaux (JLM)~\cite{JLM77} 
%recently introduced by Tostevin~\cite{Tostevin99}, 
was also tested. 
The projectile-nucleon effective interaction 
was the M3Y interaction~\cite{Bertsch77}. 
The transition density was calculated with the shell model 
using the WBT interaction~\cite{Warburton92}. 
The single-particle wave function 
was generated in a harmonic oscillator well. 
The oscillator parameter 
was chosen so that the rms radius 
corresponding to 
%implied by 
the wave function 
reproduces the experimental value~\cite{Ozawa01}: 
$b$=2.07 fm for $^{19}{\rm C}$ and 1.83 fm for $^{17}{\rm C}$. 
The effect of core polarization on quadrupole transition amplitudes 
was taken into account in the isoscaler channel, 
by introducing isospin dependent polarization charges 
obtained in the FH+RPA particle-vibration model 
and parameterized in Ref.~\cite{Sagawa04}: 
$\delta_{T=0}$=0.17 for $^{19}{\rm C}$ and 0.22 for $^{17}{\rm C}$. 
Integrated DWBA cross sections are given 
in Table~\ref{tbl:resonance_parameters} 
for each transition listed. 
The $J^{\pi}$ values are discussed below. 

In Fig.~\ref{fig:neut_19c_ad_plb}, 
the DWBA predictions of the differential cross section 
leading to the 1.46 MeV state in $^{19}{\rm C}$ are shown 
for three possible ground-state configurations. 
The solid line was obtained by using the KD02 optical potential 
for the supposed $1/2^+_1$$\rightarrow$$5/2^+_2$ transition, 
while the dot-dashed line by using the JLM potential for the same transition. 
%The JLM potential 
The latter 
was specifically derived 
for the $p$+$^{19}{\rm C}$ system at $E_p$=70 MeV. 
%using the method in Ref.~\cite{Tostevin99}. 
These curves 
agree closely with each other, 
justifying an 
%extrapolation 
extrapolated use 
in mass number of the KD02 potential 
(the nominal 
%applicable 
range is $A$=27--209) down to the $A$=19 region 
even for nuclei with large neutron/proton ratios. 
%indicating that an extrapolated use in mass number $A$ of the KD02 potential 
%(the nominal applicable range is $A$=27--209) down to the $A$=19 region 
%seems to be justified even for nuclei with large neutron/proton ratios. 
We therefore adopt the KD02 potential 
below. 
%in the present analysis. 
%in the following. 
%as for the present comparison. 
%as far as the argument based on the cross section magnitude is concerned. 
%are close to each other, 
%showing small ambiguity 
%due to the choice of the optical potential. 
%showing the small ambiguity due to the choice of the optical potential. 
Dashed and dotted curves 
respectively assume $J^{\pi}_{\rm g.s.}$=$3/2^+_1$ and $5/2^+_1$, 
and the same final state. 
%and $5/2^+_2$ for the excited state. 
These curves fail to reproduce 
the magnitude of the cross section. 
%the cross section magnitude. 
Clearly, 
the data are 
%preferentially 
much better described by 
%in much better agreement with 
the solid and dot-dashed curves 
obtained with $J^{\pi}_{\rm g.s.}$=$1/2^+_1$. 
If the 1.46 MeV levels is not $5/2^+_2$, 
it might 
%would 
be identified with the $7/2^+_1$ state. 
DWBA cross sections 
exciting this $7/2^+_1$ state 
from any of the members of the low-lying triplet, 
%from either one of the ground triplet levels, 
integrated up to $\theta_{\rm c.m.}$=64$^{\circ}$, 
are, however, at most 4.2 mb (for 5/2$^+_1$$\rightarrow$$7/2^+_1$), 
only 50\% of the data. 
We thus conclude that the spin of the ground state of $^{19}{\rm C}$ 
is consistent with $J^{\pi}$=1/2$^+$~\cite{Nakamura99,Bazin95,Maddalena01}, 
and that of the 1.46 MeV state is $5/2^+$. 

In the study of the 
$^{14}{\rm C}$($^{12}{\rm C}$,$^9{\rm C}$)$^{17}{\rm C}$ 
reaction~\cite{Bohlen07}, 
a state was observed at 2.06 MeV in $^{17}{\rm C}$, 
which was assigned as either 3/2$^+_2$ or 7/2$^+_1$ 
using the results of shell model calculations. 
Although there is a slight discrepancy in the excitation energies, 
a possible counterpart of this state is the 2.20 MeV state 
%observed 
in the present $(p,p')$ study, 
for which we examine 
the two proposed $J^{\pi}$ assignments. 
%the two cases of the spin. 
Of all possible transitions connecting 
the low-lying triplet of states 
%the near-ground triplet states 
and the ($3/2^+_2$, $7/2^+_1$) state, 
it is found that only three have sizeable cross sections 
when integrated up to $\theta_{\rm c.m.}$=64$^{\circ}$: 
2.3 mb for $1/2^+_1$$\rightarrow$$3/2^+_2$, 
2.7 mb for $3/2^+_1$$\rightarrow$$7/2^+_1$, 
and 3.0 mb for $5/2^+_1$$\rightarrow$$7/2^+_1$. 
The DWBA cross sections for other transitions 
are at most 0.66 mb ($3/2^+_1$$\rightarrow$$3/2^+_2$). 
Of the three the $1/2^+_1$$\rightarrow$$3/2^+_2$ case 
is readily excluded since it has been clearly demonstrated in a $g$-factor 
measurement that $J^{\pi}_{\rm g.s.}$ of $^{17}{\rm C}$ 
is different from $1/2^+$~\cite{Ogawa02}. 
We are thus left with the $7/2^+_1$ assignment for the 2.20 MeV state. 
DWBA predictions leading to this state 
are shown in Fig.~\ref{fig:neut_17c_ad_plb} 
for three initial-state configurations: 
solid line assumes $J^{\pi}_{\rm g.s.}$=$3/2^+_1$, 
dashed one $5/2^+_1$, and dotted one $1/2^+_1$. 
The $1/2^+_1$$\rightarrow$$7/2^+_1$ assumption 
gives much lower cross sections than the data. 
Of the remaining two 
we see that 
the $3/2^+_1$ assumption for the ground state better describes the data 
by reproducing 
the slope of the angular distribution, 
%the angular distribution slope, 
although 
the $5/2^+_1$ assumption also gives a 
reasonable 
%moderate 
description of the data. 

In Fig.~\ref{fig:neut_17c_ad_plb}, 
differential cross sections leading to the 3.05 MeV state 
are also compared with DWBA predictions 
obtained by assuming different configurations for the ground state, 
and the $9/2^+_1$ state for the excited state. 
This state was populated only weakly. 
It had a sizeable cross section only at backward angles, 
where transition amplitudes 
with large angular momentum transfers are involved. 
It could be identified as the $9/2^+$ state 
reported at 3.10 MeV in the three-neutron transfer work~\cite{Bohlen07}, 
%It was 
and 
predicted at 3.01 MeV 
%with a dominant stretched three-neutron (0$d$5/2)$^3$ configuration 
in the present shell model calculations. 
%with the WBT interaction. 
The cross sections leading to this state 
are in good agreement with the solid curve 
calculated for the $3/2^+_1$$\rightarrow$$9/2^+_1$ transition, 
excluding the possibility of $J^{\pi}_{\rm g.s.}$=$5/2^+_1$ 
for $^{17}{\rm C}$. 
This further corroborates 
the earlier $3/2^+$ assignments for the ground state 
of this nucleus~\cite{Maddalena01,Ogawa02,Sauvan00-04,Pramanik03}, 
giving us confidence in the current procedure 
employing the shell model wave functions and DWBA calculations. 

The 6.13 MeV state in $^{17}{\rm C}$ 
appears to correspond to the 6.20 MeV state reported close in energy 
in the three-neutron transfer work~\cite{Bohlen07}. 
In that study 
%In the work 
it was tentatively assigned as either $5/2^+_4$ or $5/2^+_5$, 
with a preference for the latter assignment 
due to larger occupancies predicted for the 0$d$ shells. 
The $5/2^+_4$ and $5/2^+_5$ states are located 
at 6.25 and 6.81 MeV, respectively, 
in the present shell model calculations using the WBT interaction. 
DWBA cross sections leading to the $5/2^+_4$ and $5/2^+_5$ states, 
angle integrated up to $\theta_{\rm c.m.}$=64$^{\circ}$, 
are 0.94 and 0.25 mb, 
respectively. 
The experimental value of 0.8(1) mb 
favours the $5/2^+_4$ assignment for this state. 
The $5/2^+_4$ state is predicted to have a large occupancy 
of the 0$d$3/2 shell of 23\%, 
which corresponds to almost one neutron in this shell, 
in contrast to other lower energy states involved in this study: 
4\%, 4\%, and 6\% for the ground, $7/2^+_1$, and $9/2^+_1$ states, 
respectively. 
This indicates that promoting one neutron 
from lower-lying 
%the 0$d$5/2 and 1$s$1/2 
shells 
to the 0$d$3/2 shell 
is the dominant excitation process of this state. 
The deduced narrow width is consistent 
with the reported value of $\Gamma_r$=0.35(15) MeV in Ref.~\cite{Bohlen07}. 
%The candidates for this state from this study 
%are the $5/2^+_3$ (4.00 MeV) and $7/2^+_2$ (5.35 MeV) shell model states. 
%These states are predicted to have cross sections well comparable to the 
%observed one, while the cross section leading to the $9/2^+_2$ (4.72 MeV) 
%shell model state is calculated to be only 20\% of the experimental value. 

The $5/2^+_2$ shell model state at 1.72 MeV in $^{17}{\rm C}$ 
is not observed clearly 
%could hardly be observed 
presumably 
due to the large expected width~\cite{Bohlen07}. 
The angle integrated DWBA cross section leading to this state 
from the $3/2^+_1$ ground state was calculated to be 0.69 mb. 

According to the argument of Ref.~\cite{Suzuki03a} 
the spin of $J^{\pi}_{\rm g.s.}$=1/2$^+_1$ for $^{19}{\rm C}$ 
may imply a prolate intrinsic deformation for the ground state. 
In terms of the Nilsson diagram this indicates that 
four neutrons occupy the [220$\frac12$] and [211$\frac32$] orbits, 
and one neutron occupies the [211$\frac12$] orbit in the ground state, 
where [$Nn_z\Lambda \Omega$] are asymptotic quantum numbers 
referring to large prolate deformations. 
If one assumes that the observed $1/2^+$$\rightarrow$$5/2^+$ transition 
is primarily associated with the promotion 
of the least bound neutron in the [211$\frac12$] orbit 
to the [202$\frac52$] orbit, 
one would obtain a value for the quadrupole deformation parameter 
of $\beta_2$$\approx$ 0.4, 
by referring to neutron single-particle levels in a deformed Woods-Saxon 
potential~\cite{Hamamoto07}, 
and by equating the excitation energy of 1.46 MeV 
with the energy difference of the two Nilsson orbits. 
Interestingly, this value is consistent with the result of more recent 
deformed Skyrme HF calculations predicting a local prolate minimum 
with $J^{\pi}$=$1/2^+$ at $\beta_2$=0.39~\cite{Sagawa04}. 
In the HF calculation, however, the ground state was predicted 
to be oblate with $\beta_2$=$-0.36$ and $J^{\pi}$=$3/2^+$, 
and to be more bound by 2.05 MeV than the prolate minimum. 
Moreover a local oblate minimum having $J^{\pi}$=$1/2^+$ 
with $\beta_2$=$-0.35$ 
was also predicted to be almost degenerate 
with the $J^{\pi}$=$3/2^+$ ground state. 
The present cross section for the $1/2^+$$\rightarrow$$5/2^+$ transition 
observed in $^{19}{\rm C}$ 
will be useful in distinguishing between the two $1/2^+$ states 
with different signs of $\beta_2$, 
and provide 
a clue to further investigate 
%a hint on 
the persistence of the new magic number $N$=16 
proposed in oxygen isotopes~\cite{Ozawa00,Elekes07} 
down to the carbon isotopes. 
The HF model~\cite{Sagawa04} 
gives a good account of 
%correctly reproduces 
the ground state spin of $3/2^+$ for $^{17}{\rm C}$. 
It is of interest to see if it also 
accounts for other states observed in this experiment. 

In summary, 
we 
have demonstrated 
%could show 
that 
%the measurement of 
the $(p,p')$ reaction 
using the invariant mass method in inverse kinematics 
leading to unbound resonance states in the residual nucleus 
is feasible 
for structure studies even for nuclei far from stability. 
The measurements were made on $^{19,17}{\rm C}$ at 70 MeV/nucleon. 
%to clarify the structure of the ground and low-lying states 
%of these nuclei. 
%low-lying structure of these nuclei 
%including the respective ground states. 
One resonance in $^{19}{\rm C}$ and three in $^{17}{\rm C}$ 
were observed above the particle decay threshold. 
A DWBA analysis employing shell model wave functions 
and modern nucleon-nucleus optical potentials 
was used to identify the transitions observed. 
The spin-parity of the ground state of $^{19}{\rm C}$ 
was found to be consistent with $1/2^+$, 
and that of the strongly excited 1.46 MeV state 
was assigned to be $5/2^+_2$. 
For $^{17}{\rm C}$ the observed states 
corresponded well 
%had a good correspondence 
with those reported 
in a three-neutron transfer study~\cite{Bohlen07}. 
By adding information from this experiment 
spin-parity assignments of 7/2$^+_1$ for the 2.20 MeV state and 
5/2$^+_4$ for the 6.13 MeV state were made. 
%the spins of the 2.20 and 6.13 MeV states 
%were assigned to be 7/2$^+_1$ and 5/2$^+_4$, respectively. 
The spectroscopic information from this study 
will impose stringent constraints on further theoretical investigations 
of light neutron-rich nuclei in this region. 

The authors 
%gratefully 
acknowledge 
%It is a pleasure to acknowledge 
invaluable assistance of the staff of RARF during the experiment, 
particularly, Dr.~Y.~Yano, Dr.~A.~Goto, and Dr.~M.~Kase, 
and useful discussions with Professor I.~Hamamoto 
and Professor J.~A.~Tostevin. 
This work was supported, in part, 
by the Grant-in-Aid for Scientific Research 
(Nos.~15540257 and 15740145) from MEXT Japan. 

% \appendix

% \section{}
% \label{}

\end{document}